\documentclass[aps,prd,preprint,showpacs,nofootinbib,eqsecnum,amsmath,amssymb]{revtex4-1}
\usepackage[colorlinks=true,urlcolor=blue,linkcolor=blue,citecolor=blue]{hyperref}
\usepackage{amsmath}
\usepackage{amssymb}
\usepackage{epsf}
\usepackage{color}
\usepackage{dcolumn}
\usepackage{bm}
\everymath{\displaystyle}

\begin{document} 

\title{Electrodynamics in noninertial frames}


\author{\firstname{Yuri N.}~\surname{Obukhov}}
\email{obukhov@ibrae.ac.ru}
\affiliation{Theoretical Physics Laboratory, Nuclear Safety Institute,
Russian Academy of Sciences, B. Tulskaya 52, 115191 Moscow, Russia}



\begin{abstract}
The electromagnetic theory is considered in the framework of the generally covariant approach, that is applied to the analysis of electromagnetism in noninertial coordinate and frame systems. The special-relat\-ivistic formulation of Maxwell's electrodynamics arises in the flat Minkowski spacetime when the general coordinate transformations are restricted to a class of transformations preserving the Minkowski line element. The particular attention is paid to the analysis of the electromagnetism in the noninertial rotating reference system. For the latter case, the general stationary solution of the Maxwell equations in the absence of the electric current is constructed in terms of the two scalar functions satisfying the Poisson and the biharmonic equations with an arbitrary charge density as a matter source. The classic problem of Schiff is critically revisited.
\end{abstract}

\pacs{04.20.Cv; 03.50.De} \maketitle

\section{Introduction}

Quite paradoxically, being the oldest field theory with deep theoretical and experimental developments and with the widest technological applications, the classical electrodynamics seems to be subject to the numerous controversies in the presence of the gravitational and inertial forces. Analysing the origins of such an unsatisfactory situation, one can notice that most of issues in fact grow from the too much stress put on the special relativity theory which is too often considered as an unseparable companion of the theory of electromagnetism. By taking the special relativity as a ``basis'' of the electrodynamics, one then faces a problem of ``generalizing'' the Maxwell theory from the flat Minkowski spacetime to a curved manifold, and this unfortunately leads to the puzzling inconsistencies in the zoo of ``generalizations'' and incompatibilities of their physical predictions. Here a completely different approach is pursued. Namely, we consider the generally covariant formulation of the classical electrodynamics \cite{Birk}, dismissing the role of special relativity to just a historic oddity.

In the previous literature, \cite{Schiff,Gron,Webster:1963,Webster:1973,Hones,Volkov,Heer:1968,Mo:1070,EPostB,Nikver1,Nikver2}, considerable attention was attracted to the study of electrodynamics in the noninertial reference systems, and, in particular, to the effects of rotation. In order to highlight the level of complexity of the corresponding discussions, it is worthwhile to quote from the classic textbooks by Synge \cite{Synge:1965}, who opted to ``make no attempt here to discuss the far more difficult problem of electromagnetism in a body in accelerated motion, e.g. in rotation'', and by Sommerfeld \cite{Sommerfeld} who described some apparently contradictory consequences of Maxwell's equations, pointing, however, that this ``is no objection to Minkowski's theory of moving media, which is based on the Lorentz transformation of uniform translation, but merely an indication that it is not directly applicable to problems involving rotation.'' In one of the most well known papers \cite{Schiff}, Schiff went quite far to propose a certain generalization of Maxwell's field equations by introducing fictitious charge and current densities, and according to \cite{Webster:1973}, he ``regarded as the chief value of his work, the warning it should give anyone to avoid the use of rotating coordinate axes unless he could be sure that the only field equations involved in his problem were the two homogeneous ones, that carry over to these coordinates without any change of form.''

The main aim of our paper is to demonstrate that no difficulties arise if we consistently treat the classical electrodynamics as a {\it generally covariant} theory. This formulation is based on the fundamental metric-free (or premetric) approach developed \cite{Birk} along the lines proposed by Kottler, Cartan and van Dantzig \cite{Kottler,Cartan,Dantzig}. The crucial advantage of this approach is the general covariance (inherent in the exterior calculus) of the field equations that have the same form independently of the choice of the local coordinates or frames. There is no need to ``generalize'' Maxwell's equations, or to use ``fictitious'' charges, or to guess a ``reasonable'' constitutive relation in a non-inertial reference system or in the presence of gravity. 

Another aim of this paper is to clarify the definition of the electric and magnetic fields and excitations. The fundamental objects are the Maxwell 2-form of the electromagnetic field strength $F$ and the 2-form of the electromagnetic excitation $H$. With an account of the $(1 + 3)$ decomposition of the spacetime manifold $M$, we introduce the local coordinates $x^i$, $i = 0,1,2,3$, which makes it possible to split $F$ into the electric and magnetic fields $\bm{E}, \bm{B}$, and similarly, to split $H$ into the electric and magnetic excitations $\bm{D}, \bm{H}$. The terminology goes back to Mie \cite{Mie} and Sommerfeld \cite{Sommerfeld}. It is important to stress, however, that the local coordinates are mathematical parameters which label the points of the spacetime manifold $M$, and therefore $\bm{E}, \bm{B}$ and $\bm{D}, \bm{H}$ do not have a direct physical meaning. In order to make measurements, an observer brings in an additional structure on $M$ by introducing the orthonormal coframe field $\vartheta^\alpha$ together with the dual frame field $e_\alpha$. This gives rise to the split of Maxwell's 2-form $F$ into the {\it physical} electric and magnetic fields $\bm{\mathfrak E}, \bm{\mathfrak B}$ and to the similar split of the 2-form $H$ into the {\it physical} electric and magnetic excitations $\bm{\mathfrak D}, \bm{\mathfrak H}$. In the flat spacetime and inertial reference system, an observer can choose the Cartesian holonomic coframe, and then the coordinate field coincides with the physical one: $\bm{E} = \bm{\mathfrak E}$, etc. However, in a non-inertial reference system and in the presence of gravity one should carefully distinguish coordinate and physical fields. 

The structure of the paper is as follows. In Sec.~\ref{Max}, the generally covariant formulation of the classical electrodynamics is given, and we demonstrate in Sec.~\ref{Lorentz} how the special relativistic formulation is recovered in flat spacetime for inertial reference frames. Electromagnetism on arbitrary curved manifolds is discussed in Sec.~\ref{Cur}, with the main focus on the clarification of the constitutive relations. The general formalism is then applied in Sec.~\ref{flatrot} to Maxwell's electrodynamics in the noninertial rotating reference system. In the absence of the electric current, the general stationary solution of the Maxwell equations is described in terms of the two scalar functions satisfying the Poisson and the biharmonic equations with an arbitrary charge density as a matter source. The two particular charge distributions of the rotating uniformly charged spherical shell and the pair of rotating concentric charged spheres (spherical capacitor) are studied in full detail, revisiting the classic problem of Schiff \cite{Schiff}. Technical details of using the Green function method are presented in Appendix \ref{Solve}. Finally, the results obtained are summarized in Sec.~\ref{Conclusion}.

Our basic conventions and notations are the same as in Refs. \cite{Birk,MAG}. In particular, the world indices are labeled by Latin letters $i,j,k,\ldots = 0,1,2,3$ (for example, the local spacetime coordinates $x^i$ and the holonomic coframe $dx^i$), whereas we reserve Greek letters for tetrad indices, $\alpha,\beta,\ldots = 0,1,2,3$ (e.g., the anholonomic coframe $\vartheta^\alpha$). In order to distinguish separate tetrad indices we put hats over them. Finally, spatial indices are denoted by Latin letters from the beginning of the alphabet, $a,b,c,\ldots = 1,2,3$. The metric of the Minkowski spacetime reads $g_{\alpha\beta} = {\rm diag}(c^2, -1, -1, -1)$, and the totally antisymmetric Levi-Civita tensor $\eta_{\alpha\beta\mu\nu}$ has the only nontrivial component $\eta_{\hat{0}\hat{1}\hat{2}\hat{3}} = c$, so that $\eta_{\hat{0}abc} = c\epsilon_{abc}$ with the three-dimensional Levi-Civita tensor $\epsilon_{abc}$. The spatial components of the tensor objects are raised and lowered with the help of the Euclidean 3-dimensional metric $\delta_{ab}$. We use the standard symbols $\wedge$ and $^\ast$ for the exterior product and the Hodge duality operator, respectively.

\section{Maxwell electrodynamics: general framework}\label{Max}

The generally covariant formulation of the classical electrodynamics, which is valid on an arbitrary manifold $M$ for all coordinates and reference frames, is summarised in the system \cite{Birk}:
\begin{eqnarray}
dF &=& 0,\label{maxF0}\\
dH &=& J,\label{maxH0}\\
H &=& H(F),\label{const0}
\end{eqnarray}
which encompasses the homogeneous (\ref{maxF0}) and inhomogeneous (\ref{maxH0}) Maxwell equations and the constitutive relation (\ref{const0}) between the electromagnetic field strength 2-form $F$ and the 2-form $H$ of the electromagnetic excitation. The current 3-form $J$ describes the distribution of the electric charges and currents which are the sources of the electromagnetic field. The current is obviously conserved, $dJ = 0$.

\subsection{Electric and magnetic fields, and homogeneous Maxwell equation}

Introducing the local spacetime coordinates $x^i = (t, x^a)$ on $M$, one can decompose the electromagnetic field strength 2-form $F = {\frac 12}F_{ij}dx^i\wedge dx^j$ into the electric and magnetic fields
\begin{equation}
\bm{E}\quad {\rm and}\quad \bm{B}.\label{EB0}
\end{equation}
The components ${E}_a$ and ${B}^a$ are constructed from the Maxwell tensor $F_{ij}$ as 
\begin{equation}
{E}_a = F_{a0},\quad {B}^1 = F_{23},\quad {B}^2 = F_{31},
\quad {B}^3 = F_{12},\label{EBa0}
\end{equation}
and so the electromagnetic field strength 2-form reads
\begin{eqnarray}
F &=& -\,dt\wedge {E}_adx^a + {B}^1\,dx^2\wedge dx^3 
+ \,{B}^2\,dx^3\wedge dx^1 + {B}^3\,dx^1\wedge dx^2.\label{F0}
\end{eqnarray}

It is important to note that the local spacetime coordinates $x^i = (t, x^a)$ are absolutely arbitrary -- not necessarily the Cartesian ones. If we change the local coordinates
\begin{equation}\label{coordST}
x^i\longrightarrow x^i = x^i(x'^j) \qquad \begin{cases} t = t(t',x'^a)\\ 
x^a = x^a(t', x'^b)\end{cases},
\end{equation}
the electric and magnetic fields transform into
\begin{eqnarray}
{E}'_a &=& L^b{}_a{E}_b - P_{ab}{B}^b,\label{Enew}\\
{B}'^a &=& -\,Q^{ab}{E}_b + M^a{}_b{B}^b,\label{Bnew}
\end{eqnarray}
where the transformation matrices read
\begin{eqnarray}\label{LP}
L^b{}_a = {\frac {\partial x^b}{\partial x'^a}}{\frac {\partial t}{\partial t'}} -
{\frac {\partial x^b}{\partial t'}}{\frac {\partial t}{\partial x'^a}},\quad P_{ab} = 
\epsilon_{bcd}{\frac {\partial x^c}{\partial t'}}{\frac {\partial x^d}{\partial x'^a}},\\
Q^{ab} = \epsilon^{acd}{\frac {\partial t}{\partial x'^c}}{\frac {\partial x^b}{\partial x'^d}},
\quad M^a{}_b = \left(\det {\frac {\partial x^c}{\partial x'^d}}\right) 
{\frac {\partial x'^a}{\partial x^b}}.\label{QM}
\end{eqnarray}
As we see, for a {\it general coordinate transformation} (\ref{coordST}), the components of electric and magnetic fields are mixed up. In a special case of a {\it pure spatial} transformation,
\begin{equation}
t = t',\quad  x^a = x^a(x'^b),\label{coordS}
\end{equation}
the above formulas reduce to
\begin{eqnarray}
{E}'_a &=& {\frac {\partial x^b}{\partial x'^a}}\,{E}_b,\label{Ea}\\
{B}'^a &=& \left(\det {\frac {\partial x^c}{\partial x'^d}}\right) 
{\frac {\partial x'^a}{\partial x^b}}\,{B}^b.\label{Ba}
\end{eqnarray}
In other words, the electric and magnetic fields transform contragradiently, and from the 3-dimensional point of view, $\bm{E}$ is a 3-covector, whereas $\bm{B}$ is a 3-vector. This explains the different position of indices ${E}_a$ vs. ${B}^a$. Moreover, according to (\ref{Ba}), the magnetic field is a vector {\it density} and not a true vector. 

It is straightforward to recast the homogeneous Max\-well equation into an equivalent form for 3-component variables. Substituting (\ref{F0}) into (\ref{maxF0}), we get
\begin{equation}
\bm{\nabla}\times \bm{E} + \dot{\bm{B}} = 0,
\qquad \bm{\nabla}\cdot\bm{B} = 0.\label{maxF1}
\end{equation}
Here the dot denotes the time derivative, $\dot{} = \partial_t$, and the differential nabla operator has the usual form $\bm{\nabla} = \{\partial_a\}$. Notice, however, that the curl operator ``$\bm{\nabla}\times$'' maps a 3-covector into a 3-vector density:
\begin{equation}
\begin{split}
(\bm{\nabla}\times \bm{E})^1 = \partial_2{E}_3 - \partial_3{E}_2,\\
(\bm{\nabla}\times \bm{E})^2 = \partial_3{E}_1 - \partial_1{E}_3,\\
(\bm{\nabla}\times \bm{E})^3 = \partial_1{E}_2 - \partial_2{E}_1.
\end{split}
\end{equation}

The homogeneous system (\ref{maxF0}) is solved identically by representing the electromagnetic field strength $F = dA$ in terms of the electromagnetic potential 1-form
\begin{equation}
A = -\,\Phi dt + A_adx^a,\label{A}
\end{equation}
so that we have explicitly for the electric and magnetic fields
\begin{equation}
\bm{E} = -\,\bm{\nabla}\Phi - \dot{\bm A},\qquad
\bm{B} = \bm{\nabla}\times\bm{A}.\label{EBA}
\end{equation}

\subsection{Electric and magnetic excitations, and inhomogeneous Maxwell equation}

In a similar way, we decompose the 2-form $H = {\frac 12}H_{ij}dx^i\wedge dx^j$ and construct the electric and magnetic excitations,
\begin{equation}
\bm{D}\quad {\rm and}\quad \bm{H},\label{DH0}
\end{equation}
from the components of the excitation tensor $H_{ij}$:
\begin{equation}
{H}_a = H_{0a},\quad {D}^1 \!= H_{23},\quad {D}^2 \!= H_{31},
\quad {D}^3 \!= H_{12}.\label{HDa0}
\end{equation}
The excitation 2-form then reads
\begin{eqnarray}
H &=& dt\wedge {H}_adx^a + {D}^1\,dx^2\wedge dx^3 
+ \,{D}^2\,dx^3\wedge dx^1 + {D}^3\,dx^1\wedge dx^2.\label{H0}
\end{eqnarray}
Under the change of the spacetime coordinates (\ref{coordST}), we find the 
transformation law
\begin{eqnarray}
{H}'_a &=& L^b{}_a{H}_b + P_{ab}{D}^b,\label{Hnew}\\
{D}'^a &=& Q^{ab}{H}_b + M^a{}_b{D}^b.\label{Dnew}
\end{eqnarray}
Accordingly, we conclude that $\bm{H}$ is a 3-covector, whereas $\bm{D}$ is a 3-vector density with respect to the spatial transformations (\ref{coordS}). 

Finally, we introduce the electric current density $\bm{J}$ and electric charge density $\rho$ by identifying the components of the current 3-form $J = {\frac 16}J_{ijk}dx^i\wedge dx^j\wedge dx^k$:
\begin{equation}
{J}^1 = -\,J_{023},\quad {J}^2 = -\,J_{031},\quad 
{J}^3 = -\,J_{012},\quad \rho = J_{123},\label{Ja0}
\end{equation}
thereby recasting the current 3-form into
\begin{eqnarray}
J &=& \rho\,dx^1\wedge dx^2\wedge dx^3 - dt\wedge ({J}^1dx^2\wedge dx^3 
+ \,{J}^2dx^3\wedge dx^1 + {J}^3dx^1\wedge dx^2).\label{J0}
\end{eqnarray}
It is straightforward to derive the corresponding transformation law 
\begin{eqnarray}
\rho' &=& \left(\det {\frac {\partial x^i}{\partial x'^j}}\right) 
\left[{\frac {\partial t'}{\partial t}}\rho + {\frac {\partial t'}{\partial x^b}}
{J}^b\right],\label{rnew}\\
{J}'^a &=& \left(\det {\frac {\partial x^i}{\partial x'^j}}\right) 
\left[{\frac {\partial x'^a}{\partial t}}\rho + {\frac {\partial x'^a}{\partial x^b}}
{J}^b\right].\label{Jnew}
\end{eqnarray}
Note that here we have the determinant of the $4\times 4$ Jacobi matrix ${\frac {\partial x^i}{\partial x'^j}}$ of the spacetime coordinate transformation, whereas in (\ref{QM}) and (\ref{Ba}) we have the determinant of the $3\times 3$ Jacobi matrix ${\frac {\partial x^c}{\partial x'^d}}$ of the spatial transformation. Under the pure spatial transformation (\ref{coordS}) we find
\begin{eqnarray}
\rho' &=& \left(\det {\frac {\partial x^c}{\partial x'^d}}\right)\rho,\label{rr}\\
{J}'^a &=& \left(\det {\frac {\partial x^c}{\partial x'^d}}\right) 
{\frac {\partial x'^a}{\partial x^b}}\,{J}^b,\label{Ja}
\end{eqnarray}
which shows that $\rho$ is a 3-scalar density, whereas $\bm{J}$ is a 3-vector density. 

Substituting (\ref{H0}) and (\ref{J0}) into (\ref{maxH0}), we recast the inhomogeneous Maxwell equation into 
\begin{equation}
\bm{\nabla}\times \bm{H} - \dot{\bm{D}} = \bm{J},
\qquad \bm{\nabla}\cdot\bm{D} = \rho.\label{maxH1}
\end{equation}

\subsection{Constitutive relation}

To make the theory predictive, the system (\ref{maxF0})-(\ref{maxH0}) should be supplemented by the constitutive relation (\ref{const0}) between the excitation $H$ and the field strength $F$.

The constitutive relation depends on the dynamical contents of the theory, and, in general, it can be nonlinear and even nonlocal. In the Maxwell-Lorentz electrodynamics, the constitutive relation in vacuum (i.e., in the absence of polarizable and magnetizable matter) is linear and local:
\begin{equation}
H = \lambda_0\,{}^*\!F,\qquad \lambda_0 = \sqrt{\frac {\varepsilon_0}{\mu_0}}.\label{const}
\end{equation}
Here $\varepsilon_0$ and $\mu_0$ are the electric and magnetic constants of the vacuum, and the star ${}^*$ denotes the Hodge duality operator determined by the spacetime metric.

Maxwell equations (\ref{maxF1}) and (\ref{maxH1}) are generally covariant and are valid always in all coordinates and reference systems. Moreover, the spacetime geometry and the metric is not specified. The geometrical structure of spacetime enters only the {\it constitutive law} that relates the field strength $F$ and the excitation $H$. We will give an explicit form of the relation (\ref{const}) in the next section~\ref{Cur}, where an arbitrary metric is described.

\subsection{Physical fields}

On an arbitrary curved spacetime manifold $M$, the local coordinates $x^i$ do not have any physical meaning, and accordingly, the electric and magnetic fields $\bm{E}$, $\bm{B}$ and excitations $\bm{D}$, $\bm{H}$ are not directly observable variables. In order to make electromagnetic measurements, one needs a physical observer who is in general moving in an arbitrary, mostly non-inertial, way. Leaving the details aside (an interested reader can learn the subject from \cite{Birk,Kurz}, and the references therein), for our current study it is sufficient to know that an observer brings into the generally covariant electromagnetic theory an additional geometrical structure: the coframe (or tetrad) $\vartheta^\alpha = e^\alpha_idx^i$ together with the dual frame $e_\alpha = e_\alpha^i\partial_i$ field. Here $e_\alpha^i$ is an inverse $4\times 4$ matrix to $e^\alpha_i$.

To deal with the physical fields, one should now expand the electromagnetic field strength and the excitation 2-forms $F = {\frac 12}F_{\alpha\beta}\vartheta^\alpha\wedge\vartheta^\beta$ and $H = {\frac 12}H_{\alpha\beta}\vartheta^\alpha\wedge\vartheta^\beta$ with respect to the anholonomic frame. Explicitly, we have 
\begin{equation}\label{FHan}
F_{\alpha\beta} = e^i_\alpha e^j_\beta F_{ij},\qquad H_{\alpha\beta} = e^i_\alpha e^j_\beta H_{ij}.
\end{equation}
Following (\ref{EBa0}) and (\ref{HDa0}), we introduce the decomposition of $F$ and $H$ into the electric and magnetic fields $\bm{\mathfrak{E}}$, $\bm{\mathfrak{B}}$ and excitations $\bm{\mathfrak{D}}$, $\bm{\mathfrak{H}}$ by means of identifications:
\begin{eqnarray}
{\mathfrak{E}}_a = F_{\widehat{a}\widehat{0}},\quad {\mathfrak{B}}^1 = F_{\widehat{2}\widehat{3}},
\quad {\mathfrak{B}}^2 = F_{\widehat{3}\widehat{1}},\quad {\mathfrak{B}}^3 = 
F_{\widehat{1}\widehat{2}},\label{EBa1}\\
{\mathfrak{H}}_a = H_{\widehat{0}\widehat{a}},\quad {\mathfrak{D}}^1 = H_{\widehat{2}\widehat{3}},
\quad {\mathfrak{D}}^2 = H_{\widehat{3}\widehat{1}},\quad {\mathfrak{D}}^3 = 
H_{\widehat{1}\widehat{2}}.\label{HDa1}
\end{eqnarray}
The hats over indices denote the anholonomic (frame) components. 

The relation (\ref{FHan}) between the physical fields (anholonomic objects) and the coordinate fields (holonomic objects) in the explicit 3-dimensional form then reads
\begin{eqnarray}
{\mathfrak{E}}_a &=& {\mathcal L}^b{}_a{E}_b - {\mathcal P}_{ab}{B}^b,\qquad 
{\mathfrak{B}}^a = -\,{\mathcal Q}^{ab}{E}_b +{\mathcal M}^a{}_b{B}^b,\label{Ban}\\
{\mathfrak{H}}_a &=& {\mathcal L}^b{}_a{H}_b + {\mathcal P}_{ab}{D}^b,\qquad 
{\mathfrak{D}}^a = {\mathcal Q}^{ab}{H}_b + {\mathcal M}^a{}_b{D}^b,\label{Dan}
\end{eqnarray}
where we have the transformation matrices 
\begin{eqnarray}\label{LPan}
{\mathcal L}^b{}_a &=& e^0_{\widehat{0}}e^b_{\widehat{a}} - e^0_{\widehat{a}}e^b_{\widehat{0}}
\,,\qquad {\mathcal P}_{ab} = e^c_{\widehat{0}}e^d_{\widehat{a}}\,\epsilon_{bcd}\,,\\
{\mathcal Q}^{ab} &=& e^0_{\widehat{c}}e^b_{\widehat{d}}\,\epsilon^{acd}\,,\qquad {\mathcal M}^a{}_b 
= {\frac 12}\epsilon^{acd}\epsilon_{bef} e^e_{\widehat{c}}e^f_{\widehat{d}}\,.\label{QMan}
\end{eqnarray}

\section{Flat space and inertial frames: back to Lorentz}\label{Lorentz}

When the spacetime $M$ is a flat manifold equipped with the Minkowski metric $g_{\alpha\beta} = {\rm diag}(c^2, -1, -1, -1)$, the constitutive relation (\ref{const}) reduces to the well known
\begin{equation}
\bm{D} = \varepsilon_0\bm{E},\qquad \bm{H} = {\frac 1{\mu_0}}\bm{B},\label{constM}
\end{equation}
and the Maxwell equations (\ref{maxF1}) and (\ref{maxH1}) are then recast into
\begin{eqnarray}
\bm{\nabla}\times \bm{E} + \dot{\bm{B}} = 0,
\qquad \bm{\nabla}\cdot\bm{B} = 0,\label{maxFM}\\
\bm{\nabla}\times \bm{B} - {\frac 1{c^2}}\dot{\bm{E}} = \mu_0\bm{J},
\qquad \bm{\nabla}\cdot\bm{E} = {\frac {\rho}{\varepsilon_0}}.\label{maxHM}
\end{eqnarray}

Although the whole formalism is, of course, still invariant under the general coordinate transformations (\ref{coordST}), it is natural to specialize to a restricted class of coordinate transformations which preserve the form of the Minkowski line element
\begin{equation}\label{invds}
c^2dt^2 - \delta_{ab}dx^adx^b = c^2dt'^2 - \delta_{ab}dx'^adx'^b.
\end{equation}
These are, by definition, the Lorentz transformations.

In particular, let us consider a special case of coordinate transformations (\ref{coordST}):
\begin{equation}\label{boost} 
\left.\begin{split}
t &= \gamma\,t' + \gamma\,{\frac {v_ax'^a}{c^2}},\\
x^a &= x'^a + (\gamma - 1)\,{\frac {v^av_bx'^b}{v^2}} + \gamma v^at'.
\end{split}\right\} 
\end{equation}
Here the spatial indices are raised and lowered with the help of the Euclidean 3-metric $\delta_{ab}$, and hence $v^2 = \delta_{ab}v^av^b$. The transformation (\ref{boost}) is commonly known as a Lorentz boost determined by the three constant parameters $\bm{v} = \{v^a\}$ which are physically interpreted as components of the relative 3-velocity of the two inertial systems. The Lorentz factor is defined as usual by
\begin{equation}\label{Lor}
\gamma = {\frac 1{\sqrt{1 - {\frac {v^2}{c^2}}}}}\,.
\end{equation}
One can check that (\ref{boost}) is indeed the Lorentz transformation which leaves the Minkowski line element (\ref{invds}) invariant. 

Substituting (\ref{boost}) into (\ref{LP})-(\ref{QM}), we derive
\begin{eqnarray}\label{LPL}
L^b{}_a &=& M^b{}_a = \gamma\delta^b{}_a - {\frac {\gamma^2v^bv_a}{(\gamma + 1)\,c^2}},\\
P_{ab} &=& \gamma\epsilon_{abc}\,v^c,\qquad Q^{ab} = -\,\gamma\epsilon^{abc}\,v_c/c^2,\label{QML}
\end{eqnarray}
and therefore the general coordinate transformation (\ref{Enew})-(\ref{Bnew}) reduces to
\begin{eqnarray}
\bm{E}' &=& \gamma\left(\bm{E} + \bm{v}\times\bm{B}\right) 
- {\frac {\gamma^2\bm{v}\,(\bm{v}\cdot\bm{E})}{(\gamma + 1)\,c^2}},\label{EL}\\
\bm{B}' &=& \gamma\Bigl(\bm{B} - {\frac {\bm{v}\times\bm{E}}{c^2}}\Bigr) 
- {\frac {\gamma^2\bm{v}\,(\bm{v}\cdot\bm{B})}{(\gamma + 1)\,c^2}}.\label{BL}
\end{eqnarray}
This is the usual Lorentz transformation of the electromagnetic field. 

Summarising, the special relativity is indeed recovered in the flat spacetime for inertial frames. However, this also shows that it makes no sense to view the special relativity as a starting point for the discussion of the gravitational and inertial effects in electrodynamics.

\section{Electrodynamics on a curved spacetime}\label{Cur}

As we already stressed, the Maxwell equations on curved manifolds always have the same form (\ref{maxF1}) and (\ref{maxH1}), irrespectively how strong the gravitational and inertial fields are.

The influence of the gravity and inertia is encoded in the spacetime metric that enters the Maxwell-Lorentz constitutive law (\ref{const}). In order to clarify the structure of the latter, we need a convenient parametrization of the metric. Given the local coordinates $x^i = (t,x^a)$ on the four-dimensional curved manifold $M$, we write down the spacetime interval 
\begin{equation}\label{LT}
ds^2 = V^2c^2dt^2 - \underline{g}{}_{ab}\,(dx^a - K^acdt)\,(dx^b - K^bcdt).
\end{equation}
This is the well-known Arnowitt-Deser-Misner (ADM) parametrization of the metric \cite{ADM}. Here we assume that ten functions $V = V(x^i)$, $K^a = K^a(x^i)$, and $\underline{g}{}_{ab}(x^i)$ may depend arbitrarily on the local coordinates $t,x^a$.

Therefore, the metric (\ref{LT}) describes an {\it arbitrary} geometry, and substituting  (\ref{LT}) into (\ref{const}) we recast the constitutive law into the set of explicit relations between the components of the electric and magnetic fields $\bm{E}, \bm{B}$ and the electric and magnetic excitations $\bm{D}, \bm{H}$:
\begin{eqnarray}
D^a &=& {\frac {\varepsilon_0w}{V}}\,\underline{g}{}^{ab}E_b
- \lambda_0{\frac {w}{V}}\,\underline{g}{}^{ad}\epsilon_{bcd}K^c\,B^b,\label{const1}\\
H_a &=& {\frac {1}{\mu_0wV}}\left\{(V^2 - K^2)\underline{g}{}_{ab} + K_aK_b\right\}B^b 
-\,\lambda_0{\frac {w}{V}}\,\epsilon_{adc}K^c\,\underline{g}{}^{db}E_b.\label{const2}
\end{eqnarray}
Here $K_a = \underline{g}{}_{ab}K^b$, so that $K^2 = K_aK^a = \underline{g}{}_{ab}K^aK^b$, $\underline{g}{}^{ab}$ is the inverse spatial metric, and $w = \sqrt{\det\,\underline{g}{}_{ab}}$.

From the point of view of physics, gravity and inertia affect the electromagnetic field as an anisotropic inhomogeneous medium \cite{Plebanski,Volkov:1971,LL,Bahram} in which the effective permittivity and permeability tensors are determined by $V$ and $\underline{g}{}_{ab}$, whereas $K^a$ gives rise to the effective magnetoelectric effects. Cross-check: When $V = 1$, $\underline{g}{}_{ab} = \delta_{ab}$ (hence, $w = 1$) and $K^a = 0$, the constitutive relation (\ref{const1})-(\ref{const2}) reduces to (\ref{constM}). 

Turning to the anholonomic formulation, we describe the coframe $\vartheta^\alpha = e^\alpha_idx^i$ in the Schwinger gauge $e_a^{\,\widehat{0}} =0$ (also $e_{\widehat{a}}^{\,0} =0), ~a=1,2,3$, by the components
\begin{equation}\label{coframe}
e_i^{\,\widehat{0}} = V\delta^{\,0}_i,\quad e_i^{\widehat{a}} =
W^{\widehat a}{}_b\left(\delta^b_i - cK^b\delta^{\,0}_i\right),\ a=1,2,3,
\end{equation}
where the $3\times 3$ matrix $W^{\widehat a}{}_b$ is defined as a square root of the spatial 3-dimensional metric, $\underline{g}{}_{ab} = \delta_{\widehat{c}\widehat{d}}W^{\widehat c}{}_a W^{\widehat d}{}_b$. Accordingly, we have $w = \det W^{\widehat c}{}_d$. Substituting (\ref{coframe}) into (\ref{Ban})-(\ref{QMan}), we find the explicit relation between the anholonomic and holonomic fields:
\begin{eqnarray}\label{EE}
\bm{\mathfrak{E}}{}_a &=& {\frac 1V}\,{W}^b{}_{\widehat a}(\bm{E} + c\bm{K}\times\bm{B})_b\,,\\
\bm{\mathfrak{B}}{}^a &=& {\frac 1{w}}\,W^{\widehat a}{}_b\,\bm{B}^b\,. \label{BB}
\end{eqnarray}
Here the $3\times 3$ matrix ${W}^b{}_{\widehat a}$ is inverse to ${W}^{\widehat a}{}_b$, and the vector product is defined by $\{\bm A\times\bm B\}_a=\epsilon_{abc}A^bB^c$ for any 3-vectors $A^b$ and $B^c$.

It is important to notice that the constitutive relation (\ref{const1})-(\ref{const2}) has a more compact and transparent form when it is formulated in terms of the anholonomic fields:
\begin{eqnarray}
D^a &=& \varepsilon_0\,w\,W^a{}_{\widehat b}\,\mathfrak{E}^b,\label{const1a}\\
H_a &=& {\frac {1}{\mu_0}}\,VW^{\widehat b}{}_a\,\mathfrak{B}_b
-\,\lambda_0\,w\,\epsilon_{abc}\,W^b{}_{\widehat d}\,\mathfrak{E}^d\,K^c\,.\label{const2a}
\end{eqnarray}

\section{Electromagnetism in noninertial frames: effects of rotation}\label{flatrot}

We now apply the general formalism to the analysis of Maxwell's electrodynamics in noninertial reference systems, focusing on the case of rotating frames. The earlier research \cite{Schiff,Gron,Webster:1963,Webster:1973,Hones,Volkov,Heer:1968,Mo:1070,EPostB,Nikver1,Nikver2} will be thereby critically revisited. However, before discussing specific problems, we answer the question formulated in the title of \cite{Webster:1973}: ``Which electromagnetic equations apply in rotating coordinates?'' -- These are the Maxwell equations (\ref{maxF1}) and (\ref{maxH1}), whose form does not depend on the choice of coordinates. 

The construction of the proper reference frames and local coordinates for a noninertial observer moving in the Minkowski spacetime with nontrivial acceleration $\bm{a}$ and angular velocity $\bm{\omega}$ was thoroughly discussed by Hehl and Ni \cite{HN}. In the absence of acceleration, $\bm{a} = 0$, the spacetime geometry in a rotating reference system with the local coordinates $(t, \bm{r})$ is specified by the metric (\ref{LT}) and the coframe (\ref{coframe}), where  
\begin{equation}\label{flat}
V = 1,\qquad W^{\widehat a}{}_b = \delta^a_b,\qquad
\bm{K} = -\,{\frac {\bm{\omega}\times\bm{r}}{c}}.
\end{equation}
Then the constitutive relations (\ref{const1a}) and (\ref{const2a}) are reduced to
\begin{eqnarray}\label{constMR}
\bm{D} = \varepsilon_0\,\bm{\mathfrak{E}},\qquad
\bm{H} = {\frac {1}{\mu_0}}\,\bm{\mathfrak{B}} - \lambda_0\,\bm{\mathfrak{E}}\times\bm{K}\,,
\end{eqnarray}
whereas the relations (\ref{EE}) and (\ref{BB}) are simplified to
\begin{eqnarray}
\bm{\mathfrak{E}} = \bm{E} + c\bm{K}\times\bm{B},\qquad
\bm{\mathfrak{B}} = \bm{B}\,. \label{EBM}
\end{eqnarray}

Let us assume that there is no electric current $\bm{J} = 0$, and specialize to the case of stationary fields, so that all the partial derivatives with respect to the time $t$ vanish: $\dot{\bm{B}} = 0$ and $\dot{\bm{D}} = 0$, etc. In other words, the matter source is described only by the stationary charge density $\rho(\bm{r})$.

Under these assumptions, we can solve the inhomogeneous Maxwell equations (\ref{maxH1}) by introducing the magnetic $\psi$ and electric $\varphi$ potentials for the magnetic and electric excitations:
\begin{eqnarray}
\bm{H} = \lambda_0\,\bm{\nabla}\psi,\qquad \bm{D} = -\,\bm{\nabla}\varphi.\label{HDpot}
\end{eqnarray}
The constant factor $\lambda_0$ is introduced for convenience from the dimensional reasons. With the help of the ansatz (\ref{HDpot}), the first equation (\ref{maxH1}) is trivially satisfied $\bm{\nabla}\times\bm{H} = 0$, while the second Maxwell equation (\ref{maxH1}) reduces to the Poisson equation for the electric potential
\begin{equation}
\Delta\,\varphi = -\,\rho.\label{Gauss}
\end{equation}
By combining (\ref{HDpot}) and (\ref{constMR}), we derive the physical electric and magnetic fields
\begin{eqnarray}
\bm{\mathfrak{E}} &=& -\,{\frac 1{\varepsilon_0}}\,\bm{\nabla}\varphi,\label{EM}\\
\bm{\mathfrak{B}} &=& {\frac {1}{c}}\left(\bm{\nabla}\psi
+ \bm{\mathfrak{E}}\times\bm{K}\right).\label{BM}
\end{eqnarray}
Since from (\ref{EBM}) we have $\bm{B} = \bm{\mathfrak{B}}$, then by taking a divergence of (\ref{BM}), and using the Maxwell equation (\ref{maxF1}), $\bm{\nabla}\cdot\bm{B} = 0$, and noticing that $\bm{\nabla}\times\bm{\mathfrak{E}} = 0$ in view of (\ref{EM}), we derive 
\begin{equation}
\Delta\psi - \bm{\mathfrak{E}}\cdot(\bm{\nabla}\times\bm{K}) = 0.\label{BMdiv}
\end{equation}
By making use (\ref{flat}), we find $\bm{\nabla}\times\bm{K} = -\,2\bm{\omega}/c$, and thus finally (\ref{BMdiv}) is recast into a Poisson equation for the magnetic potential
\begin{equation}
\Delta\psi = -\,{\frac {2\bm{\mathfrak{E}}\cdot\bm{\omega}}{c}}.\label{Dpsi}
\end{equation}
The solution is straightforwardly constructed with the help of an ansatz
\begin{equation}\label{psipsi}
\psi = {\frac {2}{\varepsilon_0 c}}\,\bm{\omega}\cdot\bm{\nabla}\Psi,
\end{equation}
where the new potential $\Psi$ satisfies the biharmonic equation
\begin{equation}
\Delta^2\Psi = -\,\rho.\label{bi}
\end{equation}
    
We thus have a complete set of equations (\ref{Gauss})-(\ref{bi}), that determines the physical fields: For any distribution of the charges $\rho(\bm{r})$ we just need to solve the inhomogeneous equations (\ref{Gauss}) and (\ref{bi}). The most convenient way is to use Green's function method \cite{Masud,Thir} which yields the general result (assuming that fields are vanishing at a spatial infinity):
\begin{eqnarray}
\varphi(\bm{r}) &=& \int\,d^3\bm{r}'\,G_{\Delta}(\bm{r}, \bm{r}')\,\rho(\bm{r}'),\label{phiG}\\
\Psi(\bm{r}) &=& \int\,d^3\bm{r}'\,G_{\Delta^2}(\bm{r}, \bm{r}')\,\rho(\bm{r}'),\label{psiG}
\end{eqnarray}
where the Green functions read explicitly
\begin{eqnarray}
G_{\Delta}(\bm{r}, \bm{r}') = {\frac 1{4\pi|\bm{r} - \bm{r}'|}},\qquad
G_{\Delta^2}(\bm{r}, \bm{r}') = {\frac 1{8\pi}|\bm{r} - \bm{r}'|}.\label{Green}
\end{eqnarray}
Whereas the Green function of the usual Poisson equation (\ref{Gauss}) is standard, it seems worthwhile to point readers to the reference \cite{Boyling} for the further discussion of the lesser known Green function for the biharmonic equation (\ref{bi}). 

\subsection{Rotating charged spherical shell}\label{effects1}

After reaching a complete understanding of the stationary case, and establishing the formal general solution (\ref{phiG}), (\ref{psiG}) of the problem for an arbitrary distribution of the charges $\rho(\bm{r})$, it is of interest to look more closely into the special cases. A physically important example is represented by a thin uniformly charged spherical shell (or a charged conducting solid sphere). 

For this case, when the sphere of a radius $r_0$ has the total charge $Q$, the source in the Poisson equation (\ref{Gauss}) and in the biharmonic equation (\ref{bi}) is described by the charge density
\begin{equation}
\rho(\bm{r}) = {\frac {Q}{4\pi r_0^2}}\,\delta(r - r_0).\label{rhoS}
\end{equation}
Substituting this into (\ref{phiG}), for the electric potential $\varphi$ we find (\ref{phiS}). 
Or, explicitly, 
\begin{eqnarray}
\varphi = {\frac {Q}{4\pi}}\times\left\{
\begin{split}
{\frac {1}{r}}, &  &  r > r_0,\\
{\frac {1}{r_0}}, &  &  r < r_0.
\end{split} \right.\label{phiccs}
\end{eqnarray}
This yields the electric excitation $\bm{D} = -\,\bm{\nabla}\varphi$, and hence the physical electric field (\ref{EM}):
\begin{eqnarray}
\bm{\mathfrak{E}} = {\frac {1}{4\pi\varepsilon_0}}\times\left\{
\begin{split}
{\frac {Q\,\bm{r}}{r^3}}, &  &  r > r_0,\\
0, &  &  r < r_0.
\end{split} \right.\label{Eccs}
\end{eqnarray}

Analogously, substituting the charge density (\ref{rhoS}) into (\ref{psiG}), for the magnetic potential $\Psi$ we derive (\ref{psiS}), or explicitly
\begin{eqnarray}
\Psi = {\frac {Q}{8\pi}}\times\left\{
\begin{split}
r &+ {\frac {r_0^2}{3r}}, &  r > r_0,\\
r_0 &+ {\frac {r^2}{3r_0}}, &  r < r_0.
\end{split} \right.\label{Psiccs}
\end{eqnarray}
Then by a direct differentiation we find the magnetic potential (\ref{psipsi}) 
\begin{eqnarray}
\psi = {\frac {Q\,\bm{\omega}\cdot\bm{r}}{4\pi\varepsilon_0\,c}}\times\left\{
\begin{split}
{\frac {1}{r}}\, -&\, {\frac {r_0^2}{3r^3}},   &  r > r_0,\\
{\frac {2}{3r_0}}, &   &  r < r_0.
\end{split} \right.\label{psiccs}
\end{eqnarray}
One can check that (\ref{psiccs}) solves the Poisson equation (\ref{Dpsi}) for the right-hand side with the electric field (\ref{Eccs}). 

Then, by plugging (\ref{psiccs}) and (\ref{Eccs}) into (\ref{BM}), we obtain the physical magnetic field: 
\begin{eqnarray}
\bm{\mathfrak B} = {\frac {\mu_0}{4\pi}}\times\left\{
\begin{split}
3\,{\frac {(\bm{\mathfrak m}\cdot\bm{r})\,\bm{r}}{r^5}} &- {\frac {\bm{\mathfrak m}}{r^3}}, &  r > r_0,\\
{\frac {2\,\bm{\mathfrak m}}{r_0^3}}, & &  r < r_0.
\end{split} \right.\label{IE40}
\end{eqnarray}

The final result encompasses the equations (\ref{Eccs}) and (\ref{IE40}) which give the physical electric and magnetic fields of a rotating charged sphere in the noninertial reference frame. While the electric field configuration has the same Coulomb form as for the nonrotating sphere, the magnetic field is described by the dipole configuration created by the rotation-induced magnetic moment
\begin{equation}
\bm{\mathfrak m} = {\frac {Q\,r_0^2\,\bm{\omega}}{3}}\,. \label{momsphere}
\end{equation}
It is instructive to compare this result with the magnetic field of a non-rotating uniformly magnetized solid sphere \cite{Jackson}, and with the analysis of a rotating charged spherical shell as seen in the non-rotating inertial laboratory frame \cite{Corum}. 

One can make some elementary estimates. Let us consider, for example, a charged sphere which is characterized by the parameters of a physical particle, namely, by electron's charge and mass: $Q = e$ and $m_e$. Then we can naturally assume that the radius is equal to the Compton length, $r_0 = \lambdabar = \hbar/m_ec$, and the angular velocity $\omega = m_ec^2/\hbar$. This yields a reasonable estimate ${\mathfrak m} = {\frac 23}\mu_B$ for the magnitude of the magnetic dipole moment (\ref{momsphere}) which turns out to be comparable to Bohr's magneton $\mu_B = e\hbar/2m_e$.

On the other hand, one may wonder whether such a mechanism could be relevant to the geomagnetic field, at least on the qualitative level. The modern estimates for the magnetic moment of the Earth give the value $7.72\times 10^{22}\,$A\,m$^2$. Making use of (\ref{momsphere}), with an account of $\omega_\oplus = 7.29\times 10^{-5}\,$s$^{-1}$ and $R_\oplus = 6.378 \times 10^6\,$m, we find the charge $Q \approx 2.6\times 10^{13}\,$C$\,\approx 10^{32}\,e$. In view of the huge charge needed, this is clearly not physically feasible. 

To complete the discussion, let us find the corresponding scalar and vector potentials $(\Phi, \bm{A})$ for the coordinate electric and magnetic fields (\ref{EBA}). As a preliminary step, we notice that in the outside region, $r > r_0$, a direct computation yields for the vector product
\begin{equation}
c\,\bm{K}\times\bm{B} = {\frac {Q\,r_0^2}{4\pi\varepsilon_0\,c^2}}\,\bm{\nabla}\left[
{\frac {(\bm{\omega}\cdot\bm{r})^2 - \omega^2\,r^2}{3r^3}}\right],\label{KxBout}
\end{equation}
whereas in a similar way we find for $r < r_0$:
\begin{equation}
c\,\bm{K}\times\bm{B} = {\frac {Q\,r_0^2}{4\pi\varepsilon_0\,c^2}}\,\bm{\nabla}\left[
{\frac {(\bm{\omega}\cdot\bm{r})^2 - \omega^2\,r^2}{3r_0^3}}\right].\label{KxBin}
\end{equation}
Then substituting this into (\ref{EBM}) and making use of (\ref{phiccs}), we find the scalar potential:
\begin{eqnarray}
\Phi = {\frac {Q}{4\pi\varepsilon_0}}\times\left\{
\begin{split}
{\frac {1}{r}}\biggl(1 &-{\frac {[\bm{\omega}\times\bm{r}]^2r_0^2}{3\,c^2\,r^2}}\biggr), & r > r_0,\\
{\frac {1}{r_0}}\biggl(1 &-{\frac {[\bm{\omega}\times\bm{r}]^2}{3\,c^2}}\biggr), & r < r_0.
\end{split} \right.\label{Phi1}
\end{eqnarray}
This demonstrates that the rotational motion ``distorts'' the usual Coulomb scalar potential in a peculiar way. At the same time, directly from (\ref{IE40}) we derive the vector potential
\begin{eqnarray}
\bm{A} = {\frac {\mu_0}{4\pi}}\times\left\{
\begin{split}
{\frac {\bm{\mathfrak m}\times\bm{r}}{r^3}},&  &  r > r_0,\\
{\frac {\bm{\mathfrak m}\times\bm{r}}{r_0^3}},& &  r < r_0,
\end{split} \right.\label{AS}
\end{eqnarray}
which shows that the rotation creates the magnetic field just like the ordinary magnetic moment. As usual, of course, both potentials $(\Phi, \bm{A})$ are determined only up to a gauge transformation.

\subsection{Schiff's case: rotating spherical capacitor}\label{effects2}

We are now in a position to revisit the problem which was first considered by Schiff \cite{Schiff}. Namely, let us analyse the system of the two concentric spheres with equal and opposite charges uniformly distributed over their surfaces. The corresponding electric charge density of such a spherical capacitor is a direct generalization of (\ref{rhoS}): 
\begin{equation}\label{rhoC}
\rho(\bm{r}) = {\frac {Q}{4\pi r_1^2}}\,\delta(r - r_1) - {\frac {Q}{4\pi r_2^2}}\,\delta(r - r_2),
\end{equation}
where $\pm Q$ is the total charge of each sphere, and $r_1 < r_2$ are their radii.

In view of the linearity of the problem, we can immediately make use of (\ref{phiS}) to derive the solution of the Poisson equation (\ref{Gauss}) for the electric potential:
\begin{eqnarray}
\varphi(\bm{r}) &=& \int\,{\frac {d^3\bm{r}'}{4\pi\,|\bm{r} - \bm{r}'|}}
\left({\frac {Q\,\delta(r' - r_1)}{4\pi r_1^2}} - {\frac {Q\,\delta(r' - r_2)}{4\pi r_2^2}}
\right)\nonumber\\
&=& {\frac {Q}{8\pi}}\left({\frac {|r + r_1| - |r - r_1|}{rr_1}}
- {\frac {|r + r_2| - |r - r_2|}{rr_2}}\right).\label{phiC}
\end{eqnarray}
Explicitly, in the exterior, intermediate and interior regions, we thus have
\begin{eqnarray}
\varphi = {\frac {Q}{4\pi}}\times\left\{
\begin{split}
0 , &  & r > r_2 \\
{\frac {1}{r}} \,-\,& {\frac {1}{r_2}},   &  r_1 < r < r_2,\\
{\frac {1}{r_1}} \,-\,& {\frac {1}{r_2}},   &  r < r_1.
\end{split} \right.\label{phiCex}
\end{eqnarray}
This yields the electric excitation $\bm{D} = -\,\bm{\nabla}\varphi$, and hence the physical electric field (\ref{EM}):
\begin{eqnarray}
\bm{\mathfrak{E}} = {\frac {1}{4\pi\varepsilon_0}}\times\left\{
\begin{split}
0, &  & r > r_2 \\
{\frac {Q\,\bm{r}}{r^3}}, &  &  r_1 < r < r_2,\\
0, &  &  r < r_1.
\end{split} \right.\label{EC}
\end{eqnarray}

Analogously, substituting the charge density (\ref{rhoC}) into (\ref{psiG}), and making use of (\ref{psiS}), we derive a solution of the biharmonic equation (\ref{bi}) for the magnetic potential
\begin{eqnarray}
\Psi(\bm{r}) &=& \int\,d^3\bm{r}'\,{\frac {|\bm{r} - \bm{r}'|}{8\pi}}
\left({\frac {Q\,\delta(r' - r_1)}{4\pi r_1^2}} - {\frac {Q\,\delta(r' - r_2)}{4\pi r_2^2}}
\right)\nonumber\\
&=& {\frac {Q}{16\pi}}\left({\frac {|r + r_1|^3 - |r - r_1|^3}{3rr_1}} - 
{\frac {|r + r_2|^3 - |r - r_2|^3}{3rr_2}}\right).\label{psiC}
\end{eqnarray}
Explicitly, this reads
\begin{eqnarray}
\Psi = {\frac {Q}{8\pi}}\times\left\{
\begin{split}
{\frac {r_1^2 - r_2^2}{3r}}, &  & r > r_2 \\
r - r_2 + {\frac {r_1^2}{3r}} - {\frac {r^2}{3r_2}}, & &  r_1 < r < r_2,\\
(r_2 - r_1)\left({\frac {r^2}{3r_1r_2}} - 1\right), &  &  r < r_1,
\end{split} \right.\label{PsiCex}
\end{eqnarray}
and then by a direct differentiation we find the magnetic potential (\ref{psipsi}):
\begin{eqnarray}
\psi = {\frac {Q\,\bm{\omega}\cdot\bm{r}}{4\pi\varepsilon_0\,c}}\times\left\{
\begin{split}
{\frac {r_2^2 - r_1^2}{3r^3}}, &  & r > r_2 \\
{\frac {1}{r}} - {\frac {2}{3r_2}} - {\frac {r_1^2}{3r^3}},&   &  r_1 < r < r_2,\\
{\frac {2(r_2 - r_1)}{3r_1r_2}}, &   &  r < r_1.
\end{split} \right.\label{psiCC}
\end{eqnarray}
One can check that (\ref{psiCC}) solves the Poisson equation (\ref{Dpsi}) for the right-hand side with the electric field (\ref{EC}). 

As a result, plugging (\ref{psiCC}) and (\ref{EC}) into (\ref{BM}), the physical magnetic field is obtained: 
\begin{eqnarray}
\bm{\mathfrak B} = {\frac {\mu_0}{4\pi}}\!\!\times\!\!\left\{
\begin{split}
3{\frac {(\bm{\mathfrak m}_{1+2}\cdot\bm{r})\,\bm{r}}{r^5}}
- {\frac {\bm{\mathfrak m}_{1+2}}{r^3}}, & & r > r_2,\\
{\frac {2\,\bm{\mathfrak m}_2}{r_2^3}} + 3\,{\frac {(\bm{\mathfrak m}_1\cdot\bm{r})\,\bm{r}}{r^5}}
- {\frac {\bm{\mathfrak m}_1}{r^3}}, & & r_1 < r < r_2,\\
{\frac {2\,\bm{\mathfrak m}_1}{r_1^3}} + {\frac {2\,\bm{\mathfrak m}_2}{r_2^3}}, & & r < r_1.
\end{split} \right.\label{BC}
\end{eqnarray}
This obviously describes a field configuration created by the two magnetic dipole moments
\begin{equation}
\bm{\mathfrak m}_1 = {\frac {Q\,r_1^2\,\bm{\omega}}{3}},\qquad
\bm{\mathfrak m}_2 = {\frac {-\,Q\,r_2^2\,\bm{\omega}}{3}}.\label{mm}
\end{equation}
Their sum is denoted $\bm{\mathfrak m}_{1+2} = \bm{\mathfrak m}_1 + \bm{\mathfrak m}_2$.

Finally, by making use of the results of the previous section, (\ref{KxBout})-(\ref{AS}), we can derive the scalar and vector potentials. A direct computation yields
\begin{eqnarray}
\Phi = {\frac {Q}{4\pi\varepsilon_0}}\times\left\{
\begin{split}
{\frac {[\bm{\omega}\times\bm{r}]^2\,(r_2^2 - r_1^2)}{3\,c^2\,r^3}},&  & r > r_2 \\
{\frac {1}{r}}\Bigl(1 - {\frac {[\bm{\omega}\times\bm{r}]^2r_1^2}{3\,c^2\,r^2}}\Bigr) & & \\
-\,{\frac {1}{r_2}}\Bigl(1 - {\frac {[\bm{\omega}\times\bm{r}]^2}{3\,c^2}}\Bigr),
&  & r_1 < r < r_2,\\
{\frac {r_2 - r_1}{r_1r_2}}\Bigl(1 - {\frac {[\bm{\omega}\times\bm{r}]^2}{3\,c^2}}\Bigr),&  &  r < r_1,\end{split} \right.\label{PhiCex}
\end{eqnarray}
for the scalar potential, whereas for the vector potential we find
\begin{eqnarray}
\bm{A} = {\frac {\mu_0}{4\pi}}\times\left\{
\begin{split}
{\frac {\bm{\mathfrak m}_1\times\bm{r}}{r^3}} + {\frac {\bm{\mathfrak m}_2\times\bm{r}}{r^3}},&  &  r > r_2,\\
{\frac {\bm{\mathfrak m}_1\times\bm{r}}{r^3}} + {\frac {\bm{\mathfrak m}_2\times\bm{r}}{r_2^3}},&  &  r_1 < r < r_2,\\
{\frac {\bm{\mathfrak m}_1\times\bm{r}}{r_1^3}} + {\frac {\bm{\mathfrak m}_2\times\bm{r}}{r_2^3}},&  &  r < r_1.
\end{split} \right.\label{AC}
\end{eqnarray}

As we see, Schiff's results \cite{Schiff} cannot be confirmed by our findings. Although the physical electric field (\ref{EC}) of a system of two rotating concentric charged spheres, indeed, vanishes outside the capacitor (i.e., in the external $r > r_2$ and internal $r < r_1$ regions), and inside the capacitor ($r_1 < r < r_2$) it coincides with the nonrotating field configuration, however, the magnetic field  (\ref{BC}) is nontrivial everywhere and it has a clear dipole structure created by the two rotation-induced magnetic moments (\ref{mm}).

\section{Discussion}\label{Conclusion}

We have demonstrated that the electromagnetism in the presence of the gravitational and inertial fields is consistently described in the framework of the {\it generally covariant} approach \cite{Birk} that grows from the fundamental premetric formulation of Kottler, Cartan and van Dantzig \cite{Kottler,Cartan,Dantzig}. The Maxwell field equations (\ref{maxF1}) and (\ref{maxH1}) {\it always have the same form} (for any spacetime geometry, and independently of the choice of the local coordinates or frames) when written in terms of the electric and magnetic fields $\bm{E}, \bm{B}$ and electric an magnetic excitations $\bm{D}, \bm{H}$, which arise from the $(1+3)$ decomposition of the electromagnetic field strength 2-form $F$ and the electromagnetic excitation 2-form $H$, respectively. Accordingly, there is no need to ``generalize'' Maxwell's equations, or to use ``fictitious'' charges and currents, or to guess an ``appropriate'' constitutive relation in a non-inertial reference system or in the presence of gravity.

The covariance of Maxwell's theory (\ref{maxF1}) and (\ref{maxH1}) under the general coordinate transformations (\ref{coordS}) is guaranteed by the transformation laws of the fields (\ref{Enew}), (\ref{Bnew}), excitations (\ref{Hnew}), (\ref{Dnew}), and currents (\ref{rnew}), (\ref{Jnew}), which are fixed from the $(1+3)$ decomposition of the generally covariant $F$ and $H$. The special relativity does not play any fundamental role in this approach, but the special-relativistic formulation of Maxwell's electrodynamics does naturally arise in the flat Minkowski spacetime when the general coordinate transformations are restricted to a class of transformations preserving the Minkowski line element (\ref{invds}), with the Lorentz symmetry (\ref{boost}) and (\ref{EL}), (\ref{BL}) recovered. 

The general formalism is here applied to the special case of Maxwell's electrodynamics in the noninertial rotating reference system. We show that in the absence of the electric current, the general stationary solution of the Maxwell equations can be derived in terms of the two scalar functions --the electric $\varphi$ and magnetic $\Psi$ potentials-- which satisfy, respectively, the Poisson (\ref{Gauss}) and the biharmonic equations (\ref{bi}) for an arbitrary charge density $\rho(\bm{r})$ as a matter source. The resulting system is further analysed in detail for the two particular charge distributions (\ref{rhoS}) and (\ref{rhoC}) corresponding to the rotating uniformly charged spherical shell and to the pair of rotating concentric charged spheres (spherical capacitor), respectively. The classic problem of Schiff \cite{Schiff} is thereby revisited, and we demonstrate that an everywhere nontrivial magnetic field with a dipole structure is created by the two rotation-induced magnetic moments (\ref{mm}). 

It is worthwhile to notice that the structure of the electromagnetic vector potential (\ref{AS}) and (\ref{AC}) is qualitatively the same as the gravitomagnetic vector potential of a rotating body. In the gravitoelectromagnetic approximation \cite{Ruffini,MashGEM}, the spacetime metric (\ref{LT}) is described by $V = 1 - {\mathit \Phi}/c^2$, $\underline{g}{}_{ab} = \delta_{ab}(1 + 2{\mathit \Phi}/c^2)$, $\bm{K} = 2\bm{\mathcal A}/c^2$, and for a slowly rotating massive source one finds the Lense-Thirring metric \cite{Lense,MashLT} where the gravitoelectric scalar potential ${\mathit \Phi} = GM/r$ and the gravitomagnetic vector potential $\bm{\mathcal A} = G\bm{\mathcal J}\times\bm{r}/cr^3$ are determined by the total mass $M$ and the angular momentum $\bm{\mathcal J}$ of a source. Following the early studies of Brill and Cohen \cite{BC1,BC2,BC3,BC4} of the rotating massive shells in Einstein's general relativity, one can show \cite{Poisson,Cong} that the gravitational field of a slowly rotating spherical shell of the radius $r_0$, the total mass $M$ and the total angular momentum $\bm{\mathcal J}$ is described by the gravitomagnetic vector potential
\begin{eqnarray}
\bm{\mathcal A} = G\times\left\{
\begin{split}
{\frac {\bm{\mathcal J}\times\bm{r}}{r^3}},&  &  r > r_0,\\
{\frac {\bm{\mathcal J}\times\bm{r}}{r_0^3}},& &  r < r_0,
\end{split} \right.\label{AG}
\end{eqnarray}
which is a complete analog of the electromagnetic case (\ref{AS}) of a charged rotating spherical shell.

The research is currently in progress of a more general case when the gravitational effects are taken into consideration along with the inertial ones, thereby extending the recent results obtained in \cite{Nikver1,Nikver2}, where the solution for the rotating charged spherical shell was also reported. However, it is worthwhile to note that in contrast to the Schwinger gauge (\ref{coframe}) for the coframe, the authors of the latter references used the so called Landau-Lifshitz gauge, $e^a_{\,\widehat{0}} =0$, which unnecessarily complicates the derivation of (\ref{BM}) and (\ref{Dpsi}). Moreover, it should be stressed that the choice of tetrad's gauge is not a merely technical issue, and the physical importance of the Schwinger gauge was demonstrated in \cite{ost:2009,ost:2016}.

\section*{Acknowledgments}

I am grateful to Friedrich Hehl for the long-time collaboration and constant inspiration and encouragement in this research. Numerous fruitful discussions with Oleg Teryaev and Alexander Silenko are very much appreciated. The work was supported in part by the Russian Foundation for Basic Research (Grant No. 18-02-40056-mega).

\appendix
\section{Solving Poisson and biharmonic equations: Green's function method}\label{Solve}

Substituting the charge density (\ref{rhoS}) into (\ref{phiG}), we can evaluate the volume integral in a spherical system $\bm{r}' = (r', \theta, \phi)$, where $\theta$ is an angle between $\bm{r}$ and $\bm{r}'$, so that by the cosine theorem we have
\begin{equation}
|\bm{r} - \bm{r}'| = \sqrt{r^2 + r'^2 - 2rr'\,\cos\theta}.\label{cos}
\end{equation}
Integration is straightforward:
\begin{eqnarray}
\varphi(\bm{r}) &=& \int\,d^3\bm{r}'\,{\frac 1{4\pi\,|\bm{r} - \bm{r}'|}}
{\frac {Q\,\delta(r' - r_0)}{4\pi r_0^2}}\nonumber\\
&=& {\frac {Q}{8\pi\,r_0^2}}\int\,\sin\theta\,d\theta\,\int
\,{\frac {dr'\,r'^2\,\delta(r' - r_0)}{\sqrt{r^2 + r'^2 - 2rr'\,\cos\theta}}}\nonumber\\
&=& {\frac {Q}{8\pi}}\int\,{\frac {\sin\theta\,d\theta}{\sqrt{r^2 + r_0^2 - 2rr_0\,\cos\theta}}}
\nonumber\\
&=& {\frac {Q}{8\pi}}\,{\frac {|r + r_0| - |r - r_0|}{rr_0}} = 
{\frac {Q}{4\pi}}\,{\frac {\min (r, r_0)}{rr_0}}.\label{phiS}
\end{eqnarray}

In a similar way, substituting the charge density (\ref{rhoS}) into (\ref{psiG}), and making use of (\ref{cos}), we compute the resulting volume integral that gives a solution of the biharmonic equation
\begin{eqnarray}
\Psi(\bm{r}) &=& \int\,d^3\bm{r}'\,{\frac {|\bm{r} - \bm{r}'|}{8\pi}}
{\frac {Q\,\delta(r' - r_0)}{4\pi r_0^2}}\nonumber\\
&=& {\frac {Q}{16\pi\,r_0^2}}\int\,\sin\theta\,d\theta 
\int\,dr'\,r'^2\,\sqrt{r^2 + r'^2 - 2rr'\,\cos\theta}\,\delta(r' - r_0)\nonumber\\
&=& {\frac {Q}{16\pi}}\int\,d\theta\,\sin\theta\,\sqrt{r^2 + r_0^2 - 2rr_0\,\cos\theta}
\nonumber\\
&=& {\frac {Q}{16\pi}}\,{\frac {|r + r_0|^3 - |r - r_0|^3}{3rr_0}}\nonumber\\
&=& {\frac {Q}{8\pi}}\,
{\frac 13}\left\{2(r + r_0) + (r - r_0)^2\,{\frac {\min (r, r_0)}{rr_0}}\right\}.\label{psiS}
\end{eqnarray}

\end{document}